\documentclass{aastex62}                
\newcommand{\Sp}{{\it Spitzer\/}}

\graphicspath{{./}{figures/}}

\received{July 13, 2018}
\revised{}
\accepted{}
\submitjournal{AAS Journals}

\shorttitle{\Sp\ observations of `Oumuamua}
\shortauthors{Trilling et al.}

\begin{document}

\title{\Sp\ observations of interstellar object 1I/`Oumuamua}

\author{David E. Trilling}
\affil{Department of Physics and Astronomy \\
PO Box 6010 \\
Northern Arizona University \\
Flagstaff, AZ 86011}
\affil{Lowell Observatory \\
1400 W. Mars Hill Road \\
Flagstaff, AZ 86001}

\correspondingauthor{David E. Trilling}
\email{david.trilling@nau.edu}

\author{Michael Mommert}
\affil{Department of Physics and Astronomy \\
PO Box 6010 \\
Northern Arizona University \\
Flagstaff, AZ 86011}
\affil{Lowell Observatory \\
1400 W. Mars Hill Road \\
Flagstaff, AZ 86001}

\author{Joseph L. Hora}
\affil{Harvard-Smithsonian Center for Astrophysics \\
60 Garden St., MS-65\\
Cambridge, MA 02138}

\author{Davide Farnocchia}
\affil{Jet Propulsion Laboratory, California Institute of Technology \\
4800 Oak Grove Drive \\
Pasadena, CA 91101}

\author{Paul Chodas}
\affil{Jet Propulsion Laboratory, California Institute of Technology \\
4800 Oak Grove Drive \\
Pasadena, CA 91101}

\author{Jon Giorgini}
\affil{Jet Propulsion Laboratory, California Institute of Technology \\
4800 Oak Grove Drive \\
Pasadena, CA 91101}

\author{Howard A. Smith}
\affil{Harvard-Smithsonian Center for Astrophysics \\
60 Garden St., MS-65\\
Cambridge, MA 02138}

\author{Sean Carey}
\affil{IPAC, California Institute of Technology \\
1200 E. California Boulevard \\
Pasadena, CA 91125}

\author{Carey M. Lisse}
\affil{The Johns Hopkins University Applied Physics Laboratory \\
11100 Johns Hopkins Road \\
Laurel, MD 20723-6099}

\author{Michael Werner}
\affil{Jet Propulsion Laboratory, California Institute of Technology \\
4800 Oak Grove Drive \\
Pasadena, CA 91101}

\author{Andrew McNeill}
\affil{Department of Physics and Astronomy \\
PO Box 6010 \\
Northern Arizona University \\
Flagstaff, AZ 86011}

\author{Steven R. Chesley}
\affil{Jet Propulsion Laboratory, California Institute of Technology \\
4800 Oak Grove Drive \\
Pasadena, CA 91101}

\author{Joshua P. Emery}
\affil{Department of Earth \& Planetary Science, University of Tennessee \\
306 EPS Building, 1412 Circle Drive \\
Knoxville, TN 37996, USA}

\author{Giovanni Fazio}
\affil{Harvard-Smithsonian Center for Astrophysics \\
60 Garden St., MS-65\\
Cambridge, MA 02138}

\author{Yanga R. Fernandez}
\affil{Dept. of Physics \& Florida Space Institute \\ University of Central Florida \\
4000 Central Florida Blvd. \\
Orlando, FL 32816-2385}

\author{Alan Harris}
\affil{German Aerospace Center (DLR), Institute of Planetary Research \\
Rutherfordstrasse 2, 12489 \\
Berlin, Germany}

\author{Massimo Marengo}
\affil{Iowa State University, Department of Physics and Astronomy \\
A313E Zaffarano Hall\\ Ames, IA 50011, USA}

\author{Michael Mueller}
\affil{Kapteyn Astronomical Institute \\
Rijksuniversiteit Groningen \\
PO Box 800, 9700 AV \\
Groningen, The Netherlands}
\affil{SRON, Netherlands Institute for Space Research \\PO Box 800, 9700AV \\Groningen, The Netherlands}

\author{Alissa Roegge}
\affil{Department of Physics and Astronomy \\
PO Box 6010 \\
Northern Arizona University \\
Flagstaff, AZ 86011}

\author{Nathan Smith}
\affil{Department of Physics and Astronomy \\
PO Box 6010 \\
Northern Arizona University \\
Flagstaff, AZ 86011}

\author{H. A. Weaver}
\affil{The Johns Hopkins University Applied Physics Laboratory \\
11100 Johns Hopkins Road \\
Laurel, MD 20723-6099}

\author{Karen Meech}
\affil{Institute for Astronomy \\ 
2680 Woodlawn Drive \\
Honolulu, HI 96822}

\author{Marco Micheli}
\affil{ESA SSA-NEO Coordination Centre \\
Largo Galileo Galilei, 1 \\
00044 Frascati (RM), Italy}



\begin{abstract}
1I/`Oumuamua is the first confirmed interstellar body in our Solar System.
Here we report on observations of `Oumuamua made with the \Sp\ Space Telescope on 2017 November 21--22 (UT). We integrated for 30.2~hours at 4.5~\micron\ (IRAC channel~2).
We did not detect the object
and place an upper limit on the flux of 
0.3~$\mu$Jy (3$\sigma$).
This implies an effective spherical diameter less than
[98, 140, 440]~meters and albedo greater than
[0.2, 0.1, 0.01]
under the assumption of low, middle, or high thermal beaming parameter $\eta$, respectively.
With an aspect ratio for `Oumuamua of~6:1,
these results correspond to dimensions of
[240:40, 341:57, 1080:180]~meters, respectively.
We place upper limits on the amount of dust, CO, and CO$_2$ coming from this object that are lower
than previous results; we are unable to constrain the production of other gas species.
Both our size and outgassing limits are important because
`Oumuamua's trajectory shows non-gravitational accelerations that are
sensitive to size and mass and presumably caused by gas emission.
\added{We suggest that `Oumuamua may have experienced low-level post-perihelion
volatile emission that produced a fresh, bright, icy mantle. This model
is consistent with the expected $\eta$ value and implied high albedo value for
this solution, but,
given our strict limits on CO and CO$_2$, requires
another gas species --- probably H$_2$O ---
to explain the observed non-gravitational acceleration.}
Our results extend the mystery \replaced{about}{of} `Oumuamua's origin and evolution.
\end{abstract}

\keywords{comets: individual (1I/`Oumuamua) --- minor planets, asteroids: individual (1I/`Oumuamua) --- planetary systems}


\section{Introduction}

`Oumuamua (1I/2017~U1) was discovered on 2017~October~18. 
\replaced{Within days, it became clear that its}{One week later it was announced that `Oumuamua's} orbit was unbound
\citep{U1}
and that this was the first ever discovered interstellar body --- an object that originated outside our Solar System. 

It has long been thought that comets and asteroids exist in other planetary systems. Most current models of our own Solar System suggest that today's small bodies are leftovers from the era of planet formation
\citep[e.g.,][]{dones2015},
implying that other planetary systems also produced comet and/or asteroid populations. Until now, it has been impossible to connect our own local small body populations to the large, but unresolved, groups of comets and asteroids found in exoplanetary circumstellar disks 
\citep[e.g.,][]{lisse2007,lisse2017}.

`Oumuamua was the subject of an intense,
though brief, observing campaign
\citep{jewitt2017,qz2017,knight2017,bannister2017,meech2017,masiero2017,bolin2018,fitzsimmons2018,belton2018,fraser2018,drahus2018,micheli18}.
In summary, `Oumuamua has a red, featureless visible/near infrared spectral slope; 
no directly-detected emission of gas or dust, though activity may be required to explain the presence of non-gravitational perturbations affecting its motion; 
a very elongated shape; and an excited rotation state. 
The color, spectral slope, density, and lack of apparent activity all suggest something like a D-type (primitive) asteroid,
though the implied low-level activity points to a comet-like body.
(The shape and rotation state do not particularly imply any specific analog in our Solar System.)
Assuming the object to have asteroidal density, \citet{mcneill2018} showed that no significant cohesive strength is required for `Oumuamua to resist rotational fission, \added{but even assuming a comet-like bulk density of 0.5~g/cm$^3$
we find
that a trivial cohesive strength of only 1$\pm$1~Pa is required.}.

The existence of `Oumuamua has implications for its formation and origin and on the small body populations in other planetary systems
\citep{trilling2017,
cuk2018,
feng2018,
do2018,
raymond2018a,
raymond2018b,
zwart2018,
gaidos2018,
jackson2018,
katz2018}.
Overall, these formation models generally prefer a comet-like body for interstellar interlopers.

As part of the observational campaign carried out before `Oumuamua became too faint, we observed this body with the \Sp\ Space Telescope. \Sp\ observations offered the best possibility to determine the diameter and albedo of this object
by measuring its emitted thermal infrared radiation as our team has done for thousands of Near Earth Objects (NEOs) \citep{trilling2010,trilling2016}

Here we present the results of our \Sp\ observations. We did not convincingly detect `Oumuamua and are left with an upper limit on its flux that corresponds to an upper limit on diameter and a lower limit on albedo. In Section~\ref{observations} we present our observational approach and data reduction steps;
details of the ephemeris and uncertainty calculations; and our observational results.
In Section~\ref{modeling} we present
our thermal modeling and the resulting limits on diameter and albedo, which strongly depend on choice of model parameters.
We discuss
our model results and search for activity in Section~\ref{discussion}.

\section{Observations and results \label{observations}}

\subsection{Observations and data reduction}

Observations were obtained with \Sp/IRAC \citep{fazio04} as part of the DDT program 13249. Seven Astronomical Observing Requests (AORs) were used, six of $\sim$5~hour duration with 166$\times$100 second frames, and a final 2.9 hour \added{(clock time)}
AOR with 94$\times$100 second frames, for a total of 1090 frames and 30.2 hours on-source frame time \added{(acquired over 33~hours of clock time)}. The observations were divided in this way because of limits in the number of commands and data allowed in a single AOR. The data were taken with the ``Moving Single'' target mode with Full Array readouts, using a small cycling dither pattern. Two frames were taken at each dither position, to reduce the overhead of moving after each frame. Images were obtained in both arrays, but only the 4.5 \micron\ channel was nominally centered on the target position, since the object was expected to be brightest in that IRAC channel. With the information known at that time, we estimated that with this integration time we could achieve a 3$\sigma$ or better detection of the object if it was at its expected maximum brightness during the time of observation.

\begin{figure}
\includegraphics[width=12cm]{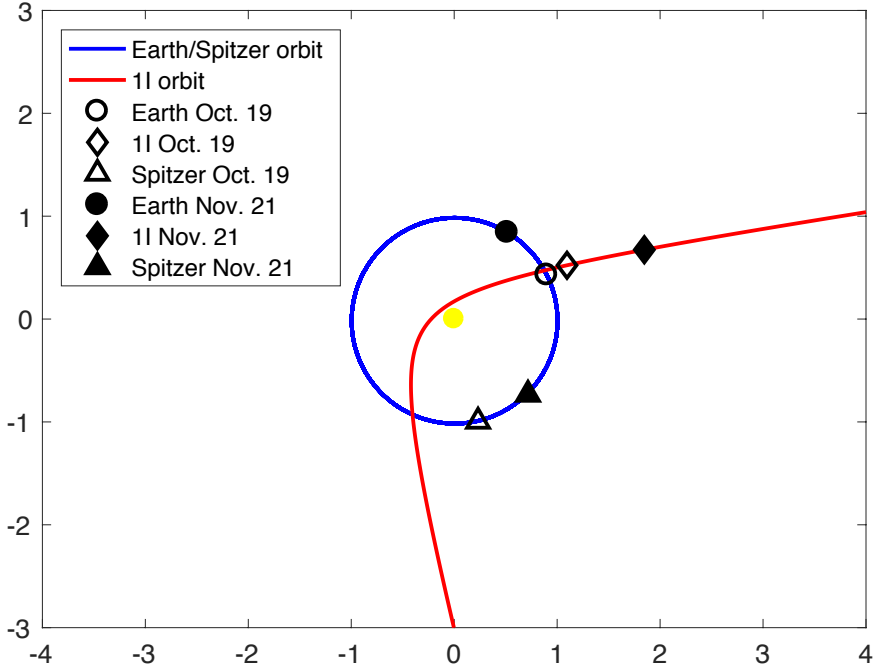}
\vspace{-15ex}
\caption{\added{Sun-centered geometry
(viewed from the north perpendicular
of the ecliptic plane)
of the Earth (circles), 
\Sp\ (triangles), and 
`Oumuamua (diamonds) at the time of 
`Oumuamua's
discovery on 2017~October~19 (open symbols)
and our 
\Sp\ observations on 2017~November~21 (filled symbols).
The \Sp\ observability window is centered on 90~degrees elongation
(that is, perpendicular to the Sun-\Sp\ line), so our \Sp\ observations 
of this object could not be executed 
until late November.
(Also, the few-week turnaround from observing request to execution was difficult
to schedule; a smaller turnaround would not have been possible.)
The orbits of Earth and Spitzer are shown as a blue (closed) circle; the 
hyperbolic orbit of `Oumuamua is shown in red.
`Oumuamua was outbound at the time of discovery.
}
\label{fig:topdown}}
\end{figure}

`Oumuamua was discovered on 2017~October~19 \added{(and identified as an interstellar body on 2017~October~25; \citet{U1})}, but because of the constraints of the \Sp\ observability zone, the earliest that the \Sp\ observations could begin was late on November 20 (Figure~\ref{fig:topdown}).
The ephemeris used to develop the original \Sp\ observation sequence was based on ground-based astrometric data through the end of October and had a prediction uncertainty larger than the \Sp\ FOV.
On November 9 the Magdalena Ridge Observatory collected additional ground-based observations, which we used together with preliminary high-precision astrometry from \citet{micheli18} to refine the orbit of `Oumuamua.
We found that the revised predicted positions could potentially put the object very close to or off the edge of the array for many frames in the AORs constructed with the original ephemeris. The \Sp\ Science Center (SSC) was able to replan the observations with the latest orbit information.
The first AOR began executing at 2017-11-21 10:13:26 UT, and the last AOR completed at 	2017-11-22 18:52:06 UT;
this is
around 2.5~months after `Oumuamua's perihelion passage.
The average heliocentric distance of `Oumuamua during the observations was 
2.0~au
and the average {\it Spitzer}-centric distance was 1.8~au; the average phase angle was around 31~degrees \added{(Figure~\ref{fig:topdown})}. This geometry changed only very slightly during the 33~hours of clock time needed to carry out these observations.
The rate of motion on-sky in these observations was around 68~arcsec/hour.

The data reduction method used was similar to that described in \citet{mommert14bd}. A mosaic of the field was constructed from the data set itself and then subtracted from the individual basic calibrated data (BCD) frames. After subtraction of the background mosaic, residual background sources and bright cosmic ray artifacts were masked in the individual BCDs before being mosaicked in the reference frame of the moving object. 

\subsection{Ephemeris and positional uncertainties}

\begin{figure}
\includegraphics[width=12cm]{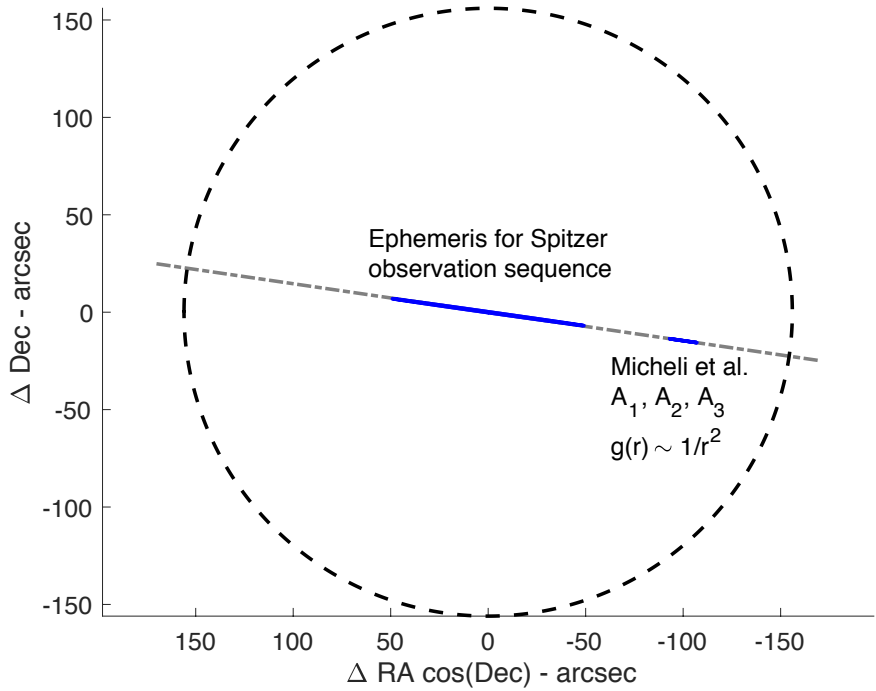}
\vspace{-15ex}
\caption{Plane-of-sky 3$\sigma$ uncertainty ellipses on 2017-11-21.0 UT for the gravity-only ephemeris used to build the Spitzer sequence (blue solid portion in the center of the frame)
and for the \citet{micheli18} solution (blue solid portion in the right of the frame),
which has radial-transverse-normal non-gravitational accelerations $(A_1, A_2, A_3) g(r)$, $g(r) \propto 1/r^2$ (where $r$ is the heliocentric distance).
The dash-dotted line is the Line of Variation, i.e., the direction of longest uncertainty. The dashed circle has a diameter of 5.2 arcmin and is therefore the inscribed circle of the {\Sp} field of view.
Thus, even when the non-gravitational accelerations that 
were not known at the time of our \Sp\ observations are included, the location of `Oumuamua is still within the IRAC field of view. 
\added{Even in the alternate and less-preferred \citet{micheli18} solutions of $1/r^k$ where $k=\{0,1,3\}$ `Oumuamua would still be in the Spitzer field of view; the uncertainty
represented by the red 
``inflated'' ellipse in Figure~\ref{mosaic} captures these other solutions.}
Furthermore,
the dithering that was used was sufficiently small that the
nominal \citet{micheli18}
location for the
object still appears in all 1,090~frames.
Note that the ellipses shown here are so narrow that they appear to be lines in this figure.
\label{fig:error_ellipses}}
\end{figure}

\citet{micheli18} later collected ground-based \added{and} Hubble Space Telescope
astrometry of `Oumuamua, eventually extending the observational arc through 2018~January~2.
Based on this longer sampling of the trajectory, they reported a 30$\sigma$
detection of a non-gravitational acceleration acting on the motion of the 
object, inferred from position measurements over time. This acceleration 
was not visible or expected when the \Sp\ observation sequence was built
in November and would have resulted in an ephemeris correction of about 
100~arcseconds at the time of the Spitzer observations
(Figure~\ref{fig:error_ellipses}).
This correction is along the Line of Variation \citep{milani05}, i.e., the direction corresponding to the semimajor axis of the uncertainty ellipse, which corresponds to a position angle (north to east) of 81.8~degrees.
Though this correction is a statistically significant deviation (7.7$\sigma$) from the gravity-only ephemeris used to build the {\Sp} observation sequence, the updated ephemeris still falls inside the {\Sp} field of view, which is
5.2~arcmin on each side.
\added{The final mosaic presented below and our data
analysis are 
based on the most recent solution for the position (i.e., \citet{micheli18}), 
so the only impact on our observations between the pre-HST solution (used for planning our
observations) and the post-HST solution would be the
error in the rate over the individual 100~second integrations. 
The difference between the two solutions (i.e., the degree of trailing
introduced)
over that length of time is completely negligible.}

\subsection{Observational results}

\begin{figure}
\includegraphics[angle=270,width=15cm]{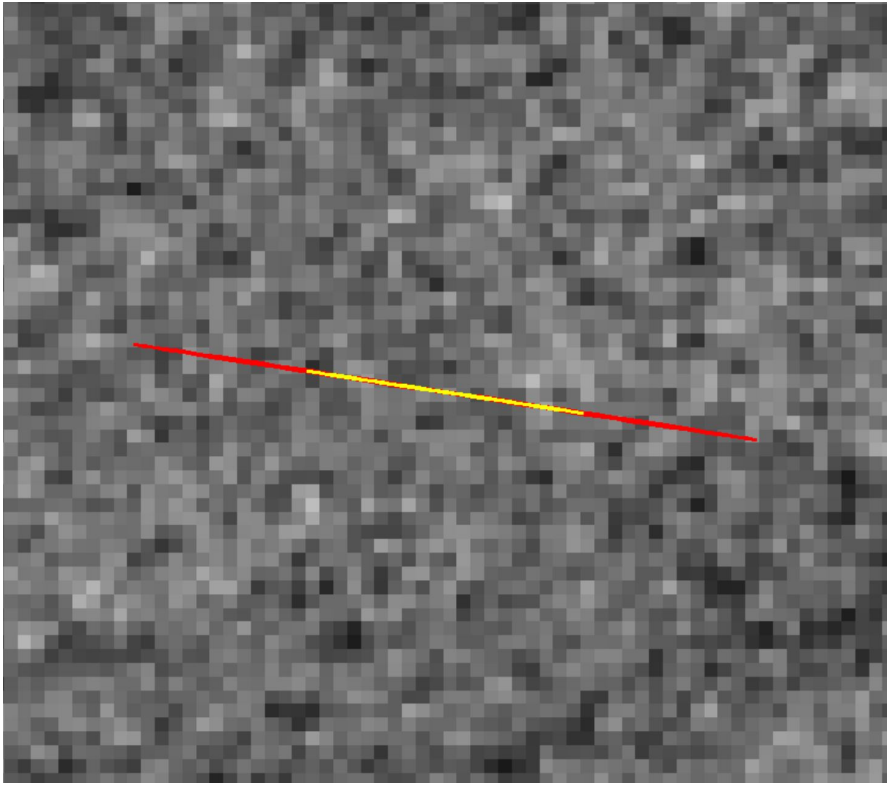}
\caption{Final combined mosaic of all IRAC Channel 2 (4.5~micron) frames in the moving frame of
`Oumuamua using zscale pixel scaling. 
Sidereal sources are removed before this stacked moving image is created, so there are no streaked stars in this image.
The yellow ellipse (which is so narrow as to appear as a line here)
indicates the
$3\sigma$ positional uncertainty at the reference time of this mosaic
as derived 
from the 
$(A_1, A_2, A_3) g(r)$, $g(r) \propto 1/r^2$ solution of \citet{micheli18}.
The red (very narrow) ellipse uses the same semi-major
axes of the yellow ellipse, 
but is inflated by a factor of three to capture ephemeris dispersions caused by the different non-gravitational models and data weight assumptions in \citet{micheli18}.
The brightest pixel
blobs that fall within these ellipses have SNR=1--2 (0.1--0.2~$\mu$Jy); the brightest blobs
in this entire image have SNR=2--3 (0.2--0.3~$\mu$Jy).
North is up and East is left;
this image has a width of 55\arcsec and a height of 50\arcsec.
The effective pixels here are 0.86~arcsec; the native pixel scale of 
IRAC is 1.21~arcsec.
}
\label{mosaic}
\end{figure}

Our final mosaic is shown in Figure~\ref{mosaic} along with the predicted location of `Oumuamua.
There are no bright coherent sources in this image, so we conclude that we did not confidently detect the source.
The 1$\sigma$ noise level in the final mosaic is $\sim$0.1~$\mu$Jy per PSF.
This noise level was determined by
recovering synthetic sources of various brightnesses that were injected in the 
final mosaic.
Sources as faint as 0.3~$\mu$Jy could be reliably found and extracted with an error of 0.1~$\mu$Jy. 
This noise floor is consistent with our expectations from \Sp\ observations of other very faint moving objects \citep{mommert14md,mommert14bd}.

There are several $\sim$2$\sigma$ ``blobs'' (essentially, single pixels) in the image, and a ``source'' that is around 1$\sigma$ that is located within the uncertainty ellipse. Since our final image is stacked in the moving frame of the target, the likelihood of any of these blobs corresponding to a true astrophysical object, which would have to be moving at the same rate as `Oumuamua over 30~hours, is vanishingly small.
Nevertheless,
the presence of these blobs in the image at the 1$\sigma$ or 2$\sigma$ level implies that there is correlated noise somewhere in our data stream. 
The final image has residuals from background stars not fully removed from the mosaics, which cause some streaking across the image. There are likely also fainter cosmic rays and other low-level array effects that survive our filtering and enter into the final image. These residuals get smeared out because of the offsets and mapping between the instrument pixels and the final image pixels on a smaller scale that will lead to correlated ``noise.'' The final image does not look like an image with only random pixel values in each pixel, but is consistent with
what we would expect for the object of interest being too faint to detect.
Alternately, this could be a tenuous detection of `Oumuamua at 1$\sigma$, or around
0.1~$\mu$Jy. In the analysis below, we use a 3$\sigma$ upper limit for our calculations, which implies
non-detection, or, at best, a weak detection.

\added{There is no significant vignetting in the 4.5~micron IRAC channel near the edge of the field. However, there could be an impact on sensitivity from the object being off the edge of the detector for some frames due to the dithering. The 3-sigma position uncertainty ellipse shown in yellow in Figure~\ref{mosaic} is fully covered by all exposures to 
within 2\% of the median coverage of the central region (small variations are caused by rejection of bad pixels or pixels affected by background objects or cosmic rays during the exposures). To the right of the yellow ellipse, the coverage drops off roughly linearly along the red path until the end where it reaches a value of around 12\% lower coverage than the central part of the image. Therefore, at that extreme end, the upper 
limit flux would then be 0.32~$\mu$Jy for objects at this position in the mosaic.
The coverage is similar along lines perpendicular to the major axes of the ellipses shown in Figure~\ref{mosaic}.
Given the small area affected and the small ($<$10\%) difference we simply use the global 3$\sigma$ (0.3~$\mu$Jy) upper limit for our analysis below.}

\section{Thermal modeling and interpretation of the non-detection \label{modeling}}

We rule out any detections of `Oumuamua at
greater than 3$\sigma$ ($\lesssim$0.3~$\mu$Jy). 
Given the geometry of the observations, we have created a model spectral energy distribution
that fits the available data: this 4.5~$\mu$m upper limit and $H_V$ (the Solar System absolute magnitude in V~band), which we take to be~22.4 \citep{meech2017}
with an uncertainty of~0.09 (using the fractional uncertainty given in \citet{meech2017}).
At 4.5~microns and 1--2~au from the Sun the flux from this object is generally dominated by thermal emission
\citep{trilling2016} \added{(modulo some low-level gas emission,
as described below)}, so a non-detection provides an upper limit on diameter and a lower limit on albedo.


We simulate the expected brightness of `Oumuamua in Spitzer IRAC
Channel 2 in order to interpret our upper limit detection. Using the
Near-Earth Asteroid Thermal Model (NEATM, \citet{harris1998}), we estimate
the target's brightness as a function of its absolute magnitude $H_V$
and a range in geometric albedo ($0.01 \leq p_V \leq 1.0$). Since the
physical properties of `Oumuamua are unknown, 
we consider a range of values for the thermal infrared beaming parameter: $\eta$=[0.8, 1.1, 2.5].
The justification for these values, which span the range of plausible $\eta$ values for almost all NEOs \citep{trilling2016} and comets \citep{fernandez2013}, is given in Section~\ref{choices}.
We account for the target's geometry at the time of our
\Sp\ observations and contributions from reflected solar light in
IRAC Channel 2
\citep{mueller2011}, assuming an infrared to optical
reflectance ratio of~1.4 \citep{trilling2016}. Furthermore, we account for color
corrections of the thermal component of the target's flux.

\begin{figure}
\includegraphics[width=15cm]{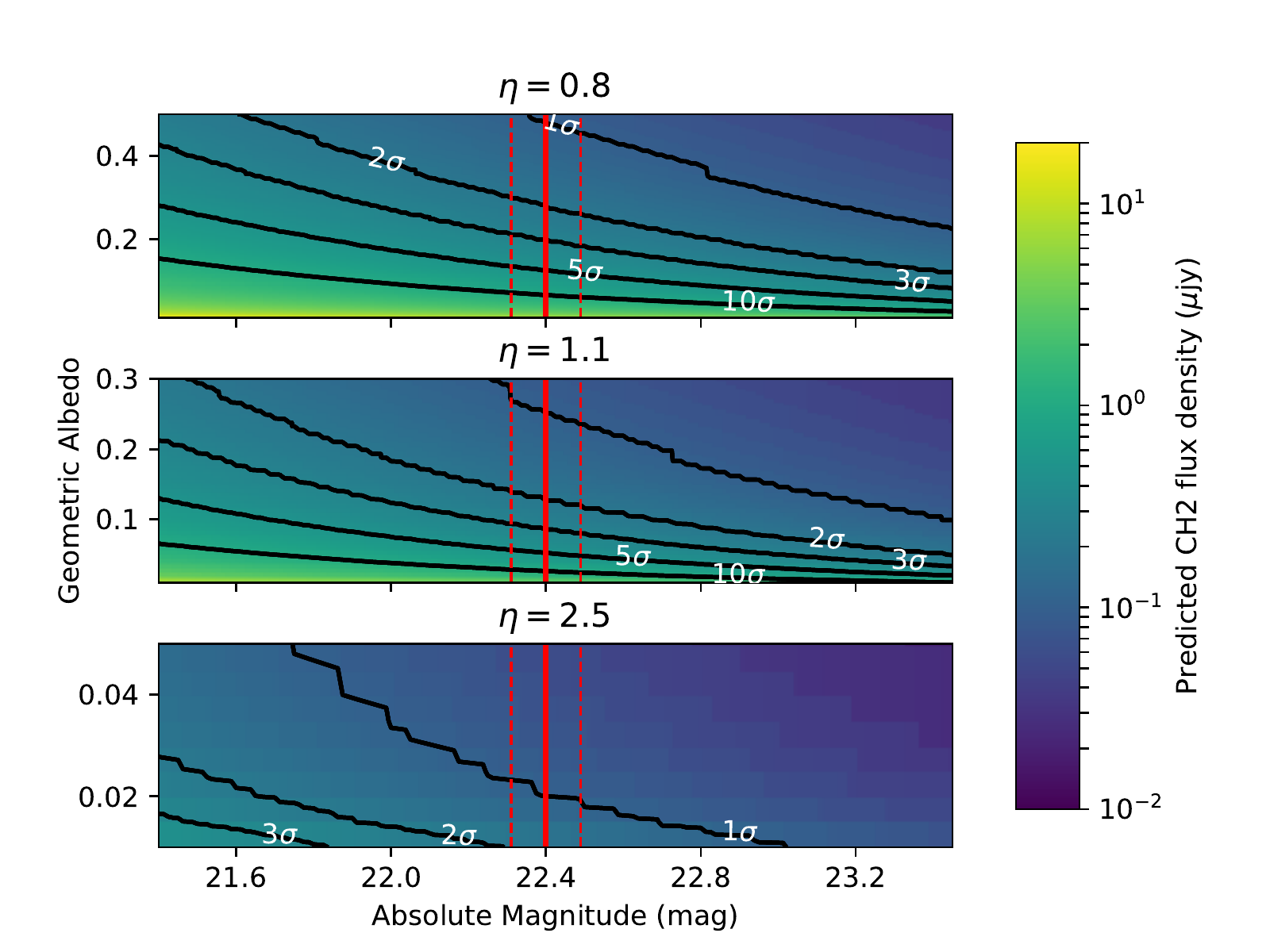}
\caption{NEATM IRAC Channel 2 (CH2) flux density prediction for `Oumuamua as a
function of the target's absolute magnitude and its geometric
albedo. The color scale represents different flux densities; black
lines indicate levels of flux density as measured from our observations.
The vertical lines indicate $H_V=22.4\pm 0.09$ (solid line and dashed lines on both sides).
Solutions for three different $\eta$ values are given, as indicated and described in the text.
Low $\eta$ requires high albedo, while $\eta=2.5$ allows any albedo.
}
\label{fig:results}
\end{figure}

\begin{figure}
\includegraphics[width=15cm]{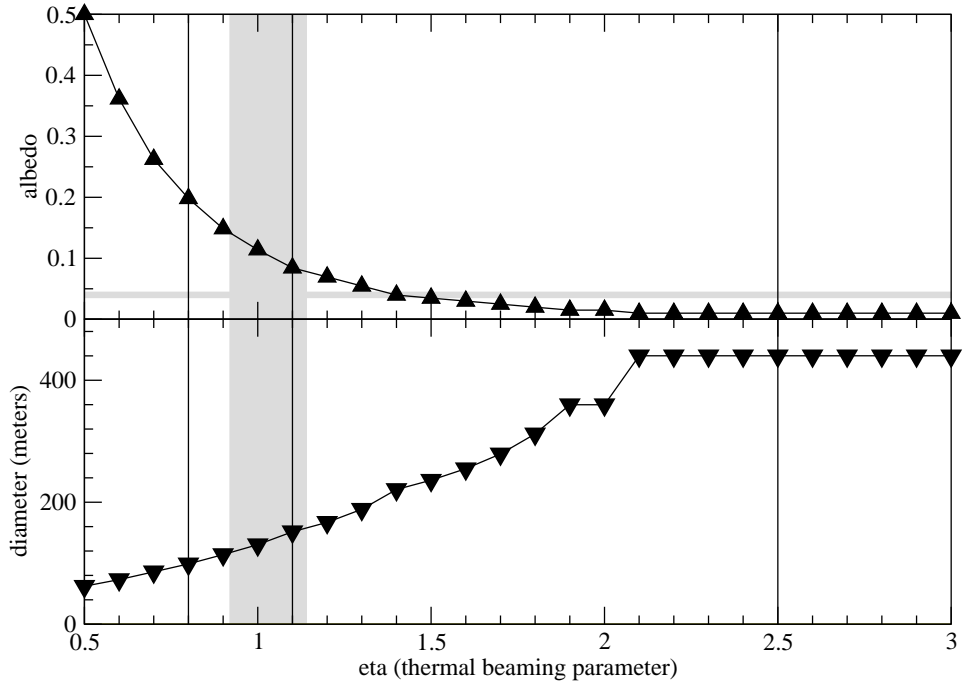}
\caption{Albedo and effective spherical diameter solutions as a function of $\eta$.
Our solutions
shown here (lower limits for albedo as upward solid triangles in the upper panel, upper limits for diameter as downward open triangles in the lower panel) correspond to the 3$\sigma$ detection thresholds described in the text and shown in Figure~\ref{fig:results}. 
\added{The three representative cases shown in Figure~\ref{fig:results} have $\eta$=[0.8, 1.1, 2.5] (thin vertical black lines). 
The typical comet albedo of~0.04, as assumed by \citet{micheli18}, is shown as the horizontal grey line in the top panel. The range of $\eta$ values for comets \citep{fernandez2013} is shown as the broad vertical grey bar in both panels. The expected result is therefore the intersection of these two grey regions, with values of $\eta \approx 1$, albedo of~0.04, and diameter around~140~meters; this result is discordant with our results.}
For $\eta > 2.0$, formally any albedo is allowed, and we choose here $p_V=0.01$ as the smallest possible value that allows a finite diameter. \added{(The non-smooth diameter function at $\eta=2$ is simply an artifact of these vanishing solutions.)}
Assuming a shape elongation of~6:1
\citep{mcneill2018}, the physical dimensions of `Oumuamua
can be calculated as 180$\times$1080~meters for the low albedo case ($\eta=2.5$) and 40$\times$240~meters for the high albedo case ($\eta=0.8$) for the short and long end-to-end dimensions, respectively.}
\label{fig:albdiam}
\end{figure}

Figure \ref{fig:results} shows the distribution of predicted IRAC Channel 2 flux
densities for the three different beaming parameters $\eta$. Black lines
indicate levels that are equal to integer multiples of the
flux density noise level measured from our data. 
For $\eta$ = [0.8, 1.1, 2.5],
the 3$\sigma$ lower limit on the target's geometric albedo is
[0.2, 0.1, 0.01], respectively (Figure~\ref{fig:albdiam}).
(Technically, any albedo is allowed for the $\eta=2.5$ case; we set here the minimum value to be~0.01 to allow for finite diameters.)
Correspondingly, we find a diameter upper limit of 
[98, 140, 440]~meters, respectively (Figure~\ref{fig:albdiam}).

\section{Discussion \label{discussion}}

\subsection{Search for activity}

Based on the discovery of non-gravitational accelerations acting upon the
orbit of `Oumuamua \citep{micheli18}, we investigate the possibility
of dust and gas activity in this object during our observations. Our
non-detection enables the placement of upper limits on the production
rates of dust, as well as CO and CO$_2$ gas; we are unable to constrain
the production of other gas species.
We use the same formalism that we used in detecting cometary behavior in the NEO Don Quixote \citep{mommert2014dq} \added{ --- \citet{bauer2015} used similar approaches with
WISE data --- }
and our measured 3$\sigma$ 4.5~micron flux density
limit (0.3~$\mu$Jy).
Within a 6~pixel (5.2~arcsec) radius
aperture (our standard size) we derive
Af$\rho \leq 0.07$~cm following the definition
of \citet{ahearn1984}, where A is albedo,
f is the filling factor, and $\rho$ is the linear radius of
the emission (here, an upper limit).
Assuming a dust particle radius of 10~$\mu$m, a 
dust bulk density of 1~g~cm$^{-3}$, an albedo of~0.03 that is
comparable with cometary dust and compatible with the range of possible
albedos that we derived for `Oumuamua, and a dust ejection velocity
equal to the expansion velocity of gas at this distance from the Sun 
\citep{ootsubo2012},
we find a 3$\sigma$ upper limit on the
dust production rate of 9~kg~s$^{-1}$. Similarly, we calculate the
3$\sigma$ upper limit on the CO$_2$ gas production rate as
9$\times$10$^{22}$~molecules~s$^{-1}$. This can be scaled into a 3$\sigma$
upper limit for the production of CO
(${\sim}$9$\times$10$^{21}$~molecules~s$^{-1}$) based on the ratio of the
CO$_2$ and CO fluorescence efficiencies
\citep{ce1983}. 

This CO upper limit is much lower than the \citet{micheli18} value of 4.5$\times 10^{25}$~molecules/s
(the most sensitive search in the literature) and implies that the outgassing from `Oumuamua cannot have CO
(or, presumably, CO$_2$) as a significant component, though the \citet{micheli18}
CO production rate assumes a relative large body and albedo of 4\%.
If `Oumuamua's size were 10--20~times smaller than the \citet{micheli18} diameter
of 220~meters then the amount of CO outgassing at the upper limit would produce sufficient
acceleration. However, an effective spherical diameter of 10--20~meters would require
an unacceptably high albedo and unacceptably low $\eta$, as described below, so this argument is rejected.
Overall, we find these upper limit production rates and our upper limit of
Af$\rho$ to be very low compared to the ensemble of comets
\citep{ahearn1995,ootsubo2012},
supporting the inactivity of `Oumuamua during our
observations.



\subsection{Uncertainties \label{uncertainties}}

Our analysis of `Oumuamua's physical properties is based on a measured
flux density upper limit and thermal modeling performed with the
NEATM. This model has been specifically designed for use on
near-Earth asteroid observations and has been shown to be reasonably accurate over a wide range of cases \citep{harris2011,mommert2018}. It is
applicable to thermal emission
from any airless body
and has been used extensively for comet nuclei as well
\citep{lisse2005,lisse2009,fernandez2013}. A
more sophisticated thermophysical model
\citep{mommert14md,mommert14bd,mommert2018}
is not appropriate here due to the lack of
information on the target (e.g., spin pole orientation and complex rotation state;
shape is somewhat known but not uniquely so) and the
upper-limit nature of the flux density measurement.

`Oumuamua is known to have a high-amplitude lightcurve 
\citep[e.g.,][]{jewitt2017,knight2017,meech2017,bolin2018,micheli18}
with, most likely, a period of 6--8~hours. Since our observations spanned 33~hours (clock time) any lightcurve effects are smoothed out and we observe only the average flux.
Furthermore, even though the \Sp\ viewing geometry of `Oumuamua is very different from that seen by observatories on and near the Earth, because `Oumuamua is in an excited rotation
state \citep{fraser2018,belton2018,drahus2018},
we likely observed the same time-averaged projected surface area that would have been seen from Earth.
\added{The lightcurves presented in \citet{belton2018} are not sinusoidal but rather have broad maxima and narrow
minima, so our 33~hour integration is likely not corrupted by faint epochs in the lightcurve.
Even in the case of a 55~hour period, one possible solution suggested by \citet{belton2018},
our observations span a significant fraction of the entire rotation and therefore included
something close to the time-averaged cross-sectional area, except in the case of a pathological orientation.}



We do not include uncertainties on the ratio of infrared to optical reflectances, as the impact of this ratio
barely affects the overall results of this study, especially in the
light of the large uncertainties on the beaming parameter $\eta$ and
hence the geometric albedo $p_V$.


We investigate the applicability of NEATM for this study given the
high aspect ratio of `Oumuamua \citep{meech2017,mcneill2018} and the
assumption of sphericity in NEATM. For this purpose, we use an
asteroid thermophysical model
\citep{mommert14bd,mommert14md,mommert2018} to derive the thermal and
reflected solar flux density of the body, \added{assuming both a highly
elongated shape and a highly oblate shape
\citep[following][]{belton2018}}. Based on \citet{mcneill2018}, we
assume a triaxial ellipsoidal shape with semi major axes~6:1:1 for the
highly elongated shape and 1:$\sqrt{6}$:$\sqrt{6}$ for the highly
oblate shape, both in arbitrary units. We furthermore use the geometry
during our \Sp\ observations, the period (7.34~hr) derived by
\citet{meech2017}, and assume a geometric albedo of~0.03 (in agreement
with our NEATM-derived lower limit) and typical small-body
values 
for thermal inertia and surface roughness. We simulate the flux
density observed at \Sp\ over one quarter of the target's rotation and
derive the average flux density, which is the quantity measured in our
observations by combining all available data. Finally, we form the
ratio of the average flux density derived for a spherical body (NEATM
assumption) to the average flux density derived from the elongated
shape and oblate shape models.  Deriving this ratio minimizes
the effects of the choice of the geometric albedo, surface roughness,
and thermal inertia used in the simulation.

We find that this flux density ratio varies as a function of the
target's spin axis latitude (as a proxy for the aspect angle of our
observations). \added{In the case of the elongated shape,} a spin axis
latitude of 90\degr (equator-on view), the ratio is~1, and rotational
effects are averaged out during our observations.  The ratio decreases
to~0.5 for spin axis latitudes approaching~0 (pole-on view), which
represents an extreme case. \added{In the case of the oblate shape, we
  find ratios 
of around [0.5, 1, 2] for $\beta$=[0, 32.7, 90]~degrees, respectively.}
As no
information on the spin axis orientation of `Oumuamua is available, we
\added{use the average\footnote{To compute this average,
we assume a 
uniform distribution of spin poles on the sphere of the body.
This is obtained by 
uniform sampling in longitude and uniform
sampling in the sine of the latitude.
Now, we compute the
average of latitude
knowing that sine of latitude is uniform.
The average from
-1 to 1 is zero (since this distribution is symmetric),
but if we take only the northern hemisphere (for example) 
then we calculate
$\int_0^1 \arcsin(x) dx$ which is 32.7\degr . We note that even under
other assumptions of the average value the deviation from our nominal solutions
are insignificant in all cases.}
latitude of 32.7\degr,
leading to a flux density ratio of~0.5 for the elongated
shape model and 0.7~for the prolate shape.}.
This mismatch between the flux densities of \added{the
different shapes} is insignificant compared to the uncertainties
introduced by the lack of knowledge of the surface properties ($\eta$)
of `Oumuamua.  We therefore conclude that our use of NEATM is
acceptable.  We also note that the uncertainties in the results from
the $H_V$ uncertainties are small compared to the uncertainties from
our lack of constraints on $\eta$.


\subsection{Discussion of possible solutions \label{choices}}

\subsubsection{Low albedo solution}

Since `Oumuamua is in an excited rotation state, absorption of solar energy could be significantly more uniform around the surface than for rotation around a single axis. This implies the temperature distribution would be smoother than a single axis rotator, requiring a higher $\eta$ than would otherwise be appropriate
\citep{myhrvold2016}.
The exact influence of the excited rotation state on the thermal emission of
`Oumuamua is difficult to model given our ignorance on its exact
rotation state and overall shape. 
The extreme of the high $\eta$ case would be represented by the Fast Rotator thermal model (FRM) \citep{mommert2018}. The FRM for this case produces virtually the same result as $\eta=2.5$ (Figure~\ref{fig:results}).
While the FRM is technically not suitable
for complex rotation, it
should be a reasonable approximation (especially since the rotation
period is not very short).

Under the conservative assumption of $\eta=2.5$ (the high $\eta$ solution) 
any albedo is allowed (Figure~\ref{fig:results} and
Figure~\ref{fig:albdiam}).
This includes arbitrarily low values. A comet-like value of~0.04
\citep{lamy2004}, as was assumed in \citet{micheli18}, implies a diameter of 220~meters, and D-type asteroids have similarly low albedos \citep{thomas2011}.
This relatively large body can still experience non-gravitational accelerations but requires relatively large impulses and, consequently, relatively high activity rates that are not commensurate with our 
CO/CO$_2$ outgassing limits presented above. 

\subsubsection{Mid-range albedo solution}

The default approach used in our Spitzer NEO 
program is to derive $\eta$ from phase angle; \citet{trilling2016}
present in some detail the correlation and dispersion in the correlation between
those two parameters.
In this case, the phase angle of 31~degrees implies $\eta$ around~1.1. This value yields $p_V>0.1$ and diameter less than 140~meters.
These values are intermediate 
in the range of acceptable solutions for `Oumuamua (Figure~\ref{fig:albdiam}).
However, even this moderate albedo is generally inconsistent with cometary albedos.

\subsubsection{High albedo solution}

Finally, a lower $\eta$ value 
appears to be more appropriate for comets \citep{fernandez2013}.
As our bounding case we take $\eta=0.8$. This implies diameter less
than 98~meters and albedo greater than~0.2
(Figure~\ref{fig:albdiam}). This small size is preferred from an
activity and non-gravitational acceleration perspective, but the high
albedo is unexpected since radiolysis of the surface during its
interstellar passage would presumably have darkened the surface 
(and reddened it; a red color is indeed observed). One possible explanation
is that `Oumuamua's recent passage by the sun was sufficient to emplace a
thin layer of bright, fresh ice on the surface, as discussed below.

If `Oumuamua has a high albedo then its inferred size (98~meter diameter) is substantially smaller than the 220~meter diameter that was assumed by \citet{micheli18}, and its mass is smaller by the cubed ratio of these solutions ($ (98/220)^3 = 1/11$).
With a smaller mass, greater acceleration is produced for a given force (i.e., outgassing).
However, force is proportional to the production rate,
and the CO production rate derived here is $10^4$~times less than that used by \citet{micheli18} to explain the measured astrometry.
This rules out the the possibility that CO or CO$_2$ outgassing was responsible for the non-gravitational acceleration that \citet{micheli18}
detected.
\added{However, we can not put constraints on outgassing of water ice, which is the other main 
volatile ice found in comets, using our data (see below).}
\deleted{
There furthermore are questions about whether a high albedo is consistent with the required mass loss, and also whether volatiles can remain in such a small body.}

\subsection{Summary of results and a possible interpretation}

\added{There are several possible interpretations of our results, as follows.
We note that in all cases, given our upper limit on CO and CO$_2$
production rates, some other gas species (e.g., water) must also have been emitted
to explain the non-gravitational acceleration observed by \citet{micheli18}.
We can place no constraint on these other
gas species.}

\added{(1)} `Oumuamua could have a high $\eta$, which would not be unusual for asteroids
but would be very unusual for comets --- although a body in an excited
rotation state might have a higher than expected $\eta$ value. In the high $\eta$ case, the albedo is
low, which means the diameter is large. However, a large body implies a large outgassing rate,
which we do not see for CO and CO$_2$ and for dust.
Conversely, \added{(2)} `Oumuamua could have a low $\eta$, in accordance with expectations for cometary
bodies. However, this requires an albedo that is much higher than that expected
for comets. This high albedo corresponds to a small diameter, which is favored,
considering our upper limits on gas and dust production.
\added{(3)} Intermediate values of 
$\eta$, albedo, and diameter are also possible, though these are not really
consistent with any expectations.

In conclusion, there is no simple asteroidal or cometary physical model that agrees with
expectations and previous work (including non-gravitational acceleration)
and our results for all of $\eta$, albedo, and diameter.
One plausible explanation is that `Oumuamua was a dormant comet nucleus reactivated,
after millions of years in interstellar space, by heating during its close passage by the Sun.
This reactivation either destroyed the thin dark mantle expected to be created by cosmic rays and galactic 
ultraviolet radiation \citep[e.g.,][]{lisse1998,lisse2004}
and/or coated the surface with an optically thick layer of new, fresh ice. In the latter case, the 
bright coating could plausibly have come from CO, CO$_2$, or water, as follows.

We assume that `Oumuamua is outgassing $9\times 10^{22}$~molecules of CO$_2$ per second (see above).
In the high albedo case, 
the effective spherical diameter of `Oumuamua is around 98~meters, and the 
surface area is therefore around $3\times 10^4$~m$^2$ (taking 49~meters as the radius of the equivalent sphere).
If we require a uniform surface layer that is 10~microns thick --- so that the surface appears bright with CO$_2$ ice for observations made at 4.5~microns 
--- then the volume of this surface layer is around 0.3~m$^3$.

\replaced{The size of a CO$_2$ molecule is around $2\times 10^{-10}$~m, 
so the volume of a single CO$_2$ molecule (assumed to be spherical) is around $3\times 10^{-29}$~m$^3$.
Therefore, the number of molecules needed is

\[
\frac{0.3~{\rm m}^3}{3\times 10^{-29}~{\rm m}^3/{\rm molecule}}
\]

\noindent which is around $10^{28}$~molecules.}
{The density of CO$_2$ ice is approximately 1.5~g/cm$^3$.
The mass required to create a surface layer of 0.3~m$^3$ is
therefore $4.5\times 10^5$~g.
We calculate the number of CO$_2$ molecules required as

\[
\frac{4.5\times 10^5~{\rm g}}{44~{\rm g/mole}} \times 6.02\times 10^{23}~{\rm  molecules/mole}
\]

which is around $6\times 10^{27}$~molecules of CO$_2$.}
At 
$9\times 10^{22}$~molecules/sec that corresponds to around 67,000~seconds, or around 0.8~days --- far less than the few weeks of `Oumuamua's perihelion passage time.
Thus, even if the efficiency of this process is small, it is still quite plausible that a low level of activity --- induced by solar heating of a near-subsurface CO$_2$ reservoir --- could produce enough material to coat the surface with bright, fresh CO$_2$ and increase the albedo to the relatively high value required in our high-albedo case.

\added{Alternately, heating of water ice into gas could present a plausible scenario.
Cometary dust:gas ratios are typically around~5:1, so our
dust emission upper limit of 9~kg/sec corresponds to 
1.8~kg/sec as an upper limit for gas emission. If we assume that all of this gas emission
is in H$_2$O, then we find 

\[
\frac{1.8\times 10^3~{\rm g/sec}}{18~{\rm g/mole}} \times 6.02\times 10^{23}~{\rm  molecules/mole}
\]

\noindent which is around $6\times 10^{25}$~molecules/sec, enough to produce the
non-gravitational accelerations reported by \citet{micheli18}.}

\added{
CO+CO$_2$ ice in typical Solar System comets is around 15\% of the water abundance.
Here our limits imply around 0.15\% for this ratio --- a factor of 100~times smaller.
This could imply that `Oumuamua was heated to $\sim$100~K prior to our observations --- either
by our Sun, or before entering our Solar System.
`Oumuamua,
if propelled by water ice sublimation at $6\times 10^{25}$~molecules/sec while
producing only
$\leq9\times 10^{22}$~molecules/sec of CO+CO$_2$,
must have been previously devolatilized of these more volatile ices.} 

\subsection{Possible analogies}

Unfortunately, we do not have pre-perihelion observations to compare to these post-perihelion observations to test the hypothesis that `Oumuamua brightened during its perihelion passage.
Further modeling of `Oumuamua's outgassing --- whether CO, CO$_2$, or some other species --- would
be very beneficial.

\added{A plausible analogy for such activity-produced resurfacing is the well-studied 
comet~67P, the target of the Rosetta mission.
\citet{keller2017} and \citet{liao2018}
showed that the nucleus of 67P
was partially resurfaced through re-condensation of volatiles released from the nucleus;
\citet{liao2018} found that the deposition rate of water ice could be up to 
several microns in an hour near perihelion.
While this does not correspond directly to
low levels of activity and an albedo increase, as proposed here for
`Oumuamua,
it is nevertheless evidence that activity can
resurface small body surfaces after perihelion passage, at the order of magnitude 
required for `Oumuamua (1--10~microns deposited in days or weeks).
\citet{bolin2018} saw no color changes as a function of rotation,
which could imply a relatively uniform resurfacing process.
This suggested emission must be
too small to create a measurable change in velocity after the first 
`Oumuamua observations (i.e., the beginning of the observational arc), or
else occurred
after perihelion but before the discovery observation of `Oumuamua (see Figure~\ref{fig:topdown}),
in order to be consistent with the results reported in \citet{micheli18}.}

\added{Another possible analogy is the well-studied comet Shoemaker-Levy~9.
\citet{sekanina1995} shows that after the breakup of this body from
tidal forces exerted by Jupiter many fragments appeared intrinsically brighter -- as if
fresh ice had just been revealed or deposited onto their surfaces.
It is possible that the shape and/or rotation state of `Oumuamua were affected
by its passage near the Sun, and interior volatiles may also have been liberated
onto the surface at the same time.
}






\section{Conclusions}

We observed interstellar body `Oumuamua for 30~hours of integration time 
at 4.5~microns with the \Sp\ Space Telescope. We did not convincingly detect the object
and place upper limits on its flux during our observations.
Depending on the assumptions used in our thermal model, we find low-, medium-, and high-albedo solutions (and corresponding limits on the effective spherical diameter).
We do not detect any activity from `Oumuamua and place upper limits for \added{CO and CO$_2$} emission that
are far lower than were derived by \citet{micheli18}
under the assumption of a body with 4\% albedo; \added{we can place no constraints on
emission of other gas species (e.g., water ice)}.
The nature of the gas emission and the origin of the non-gravitational accelerations are still unknown.

It is not clear what type of body in our Solar System is the most similar to `Oumuamua, as there are significant failures with both comets and primitive (D-type) asteroids as end-member analogs.
One possible scenario that appears to explain many of the observed
properties of `Oumuamua, including our observations, is
\added{exposure or creation, from outgassing, of a fresh, icy, bright surface
due to thermal reactivation during `Oumuamua's close perihelion passage
in September, 2017.}
However, due to the geometry of `Oumuamua's passage through the Solar System, 
there will be no more observations of this object, so it is likely that we will never know
the true nature of this interstellar interloper.

\acknowledgments
This work is based in part on observations made with the \Sp\ Space 
Telescope, which is operated by the Jet Propulsion Laboratory, 
California Institute of Technology under a contract with NASA. 

We thank the SSC Director for approving these DDT observations and the SSC staff for rapidly implementing these observations with their usual technical excellence.
Part of this research was conducted at the Jet Propulsion Laboratory, California Institute of Technology, under a contract with NASA.
KM acknowledges support from NSF awards
AST1413736 and AST1617015.

%

\vspace{5mm}
\facilities{Spitzer(IRAC)}


\software{MOPEX \citep{makovoz06}, IRACproc \citep{schuster06}}

\end{document}